\documentclass{natureprintstyle}
\usepackage{graphicx}
\usepackage{dcolumn}
\usepackage{bm}
\usepackage{amsmath}
\usepackage{amssymb}
\usepackage{float}
\usepackage{lipsum}
\usepackage{color}
\usepackage{times}
\usepackage{textcomp}
\usepackage{latexsym}
\usepackage[hidelinks]{hyperref}
\usepackage{mathtools}
\usepackage{url}

\usepackage[utf8]{inputenc}

\begin{document}

\title{Probing the existence of ultralight bosons with a single gravitational-wave measurement}

\author{Otto A. Hannuksela$^{1}$, 
Kaze W. K. Wong$^{2}$, 
Richard Brito$^{3,4}$, 
Emanuele Berti$^{2,5}$, 
Tjonnie G. F. Li$^{1}$
} 

\maketitle

\begin{affiliations}
\item 
Department of Physics, The Chinese University of Hong Kong, Shatin, N.T., Hong Kong.
\item 
Department of Physics and Astronomy, Johns Hopkins University, 3400 N. Charles Street, Baltimore, MD 21218, USA.
\item 
Max Planck Institute for Gravitational Physics (Albert Einstein Institute), Am M\"uhlenberg 1, Potsdam-Golm, 14476, Germany
\item 
Dipartimento di Fisica, ``Sapienza'' Universit\'a di Roma \& Sezione INFN Roma1, Piazzale Aldo Moro 5, 00185, Roma, Italy
\item 
Department of Physics and Astronomy, The University of Mississippi, University, MS 38677, USA
\end{affiliations}

\smallskip

\maketitle

\begin{abstract}
Light bosons, proposed as a
possible solution to various problems in fundamental physics and cosmology\cite{Bertone:2004pz,Arvanitaki:2009fg,Marsh:2015xka}, include a broad class of candidates for beyond the
Standard Model physics, such as dilatons and moduli\cite{Arvanitaki:2014faa}, wave dark matter\cite{Schive:2014dra} and axion-like particles\cite{Cardoso:2018tly}.
If light bosons
exist in nature, they will spontaneously form ``clouds'' by extracting rotational
energy from rotating massive black holes through superradiance, a
classical wave amplification process that has been studied for
decades\cite{1971JETPL..14..180Z,Press:1972zz}.
The superradiant growth of the cloud sets the geometry of the final
black hole, and the black hole geometry determines the shape of the
cloud\cite{Detweiler:1980uk,Arvanitaki:2010sy,Brito:2014wla}. 
Hence, both the black hole geometry and the cloud encode information about the light boson. 
For this reason, measurements of the gravitational field of the
black hole/cloud system (as encoded in gravitational waves) are
over-determined. 
We show that a single gravitational wave measurement can be used to
verify the existence of light bosons by model selection, rule out
alternative explanations for the signal, and measure the boson
mass. 
Such measurements can be done generically for bosons in the mass
range $[10^{-16.5},10^{-14}]$~eV using LISA observations of extreme
mass-ratio inspirals.
\end{abstract}

Gravitational waves allow us to measure to exquisite accuracy the host black hole mass and spin $(M,\,a)$, which gives us a model prediction for the boson cloud profile. 
We match this prediction with a direct measurement of the properties of the boson cloud profile (as encoded in two ``shape parameters'' $A$ and $B$, defined below) to confirm the model with no tuneable parameters. 
Such confirmation is possible when superradiant instability has occurred, and the black hole/cloud system is in equilibrium during the measurement.

Superradiance occurs when the boson Compton wavelength
$\lambda=\hbar/(m_s c)$ is comparable with the black hole's
Schwarzschild radius $R=2GM/c^2$, or $R/\lambda=0.15 ( M/10^6M_\odot )
(m_s c^2/10^{-17}{\rm eV})\sim 1$. Then the instability quickly extracts rotational energy from the black hole, leading the black hole/cloud system to equilibrium on a so-called ``Regge trajectory''\cite{Arvanitaki:2010sy}, where the rotational frequency of the boson (which from now on, for simplicity, we assume to be a scalar field) is comparable to the black hole rotational frequency\cite{Arvanitaki:2010sy,Brito:2014wla}:
\begin{equation} \label{eq:regge}
 \begin{split}
  \mu_s^{(1)} &\simeq \frac{ a}{2 M r_+},
 \end{split}
\end{equation}
where $\mu_s \equiv m_s/\hbar$ is the boson mass,
$r_+=\sqrt{M^2-a^2}+M$ is the outer horizon of the rotating black
hole, and from now on we will use geometrical units ($G=c=1$).
The superscript $(1)$ labels one of three possible
experimental ways to measure the mass [Eqs.~(\ref{eq:regge})-(\ref{eq:evolutioneq})].
To a good approximation, the scalar field profile in the equilibrium configuration is well described by\cite{Brito:2014wla,Ferreira:2017pth} 
$\varphi(t,r,\theta,\phi)=A B r e^{-Br/2}\cos(\phi-\omega_R
t)\sin\theta$.
Here $A$ is the scalar field amplitude, $B=M\mu_s^2$ is a ``scale'' parameter (note that the radial profile of the cloud has a maximum at $r_{\rm max}=2/B$), and $\omega_R\simeq \mu_s$. 
Both $A$ and $B$ are determined through independent physical processes: $A$ is set by the evolution of the black hole/cloud system, while $B$ is set by the black hole geometry when the black hole/cloud system is in equilibrium.

For typical black hole/cloud systems of interest $M\mu_s\sim 1$, so that the field oscillation time scale $\sim 1/\omega_R$ is of the order of seconds (hence much shorter than the Laser Interferometric  Space Antenna (LISA) observation time $T_{\rm obs}$) when $M\sim 10^6M_\odot$. 
Therefore we can time-average the gravitational potential generated by the cloud. 
In the equatorial plane, the result has the form $\Phi_b(r)=\Phi_b(A,\,B,\,M,\,r)$ (see Eq.~(\ref{eq:potential}) in the Supplementary Material).
By imposing that $\Phi_b(r)\sim -M_s/r$ at large $r$, where $M_s$ is the total mass in the boson cloud, we can relate the scalar field amplitude to its mass: $A=(\mu_s^2 M^{3/2} M_s^{1/2})/[8 \pi (4-\mu_s^2 M^2)]^{1/2}$.
Therefore, if we can measure the amplitude $A$ and scale $B$ of the scalar cloud we get two more estimates of the boson mass:
\begin{equation} \label{eq:shape}
 \begin{split}
  \mu_s^{(2)} &=  \left( M/ B \right)^{-1/2}
 \end{split}
\end{equation}
and
\begin{equation} \label{eq:evolutioneq}
 \begin{split}
  \mu_s^{(3)} &\simeq \ 2  \left[ \frac{\pi A^2}{M M_s} \left( \sqrt{1+\frac{2 M_s}{A^2 M \pi}}-1 \right) \right]^{1/2}.
 \end{split}
\end{equation}
To infer $\mu_s^{(3)}$, we need an estimate for the mass of the
boson cloud $M_s$. 
Although $M_s$ can be obtained from the evolution of the black
hole/cloud system\cite{Brito:2014wla} given $\mu_s^{(1)}$ measured
from Eq.~(\ref{eq:regge}), the host's initial spin and accretion rate
are unknown, and therefore $M_s$ can have any value in a range $M_s \in [0, M_{s}^{\rm max}]$. We fix $M_{s}^{\rm max}$ by assuming that the initial black hole spin (pre-superradiant amplification) is maximal. 
Even under this conservative estimate, we find that the ultralight boson hypothesis can be either confirmed or ruled out.

The superradiant instability occurs on a time scale
$\tau_{\rm inst}\sim 10^5{\rm yr}\,
j_{\rm in}^{-1}(10^6M_\odot/M)^8 (10^{-17}{\rm
      eV}/m_s c^2)^9$, where
    $j_{\rm in}$ denotes the dimensionless spin of the black hole before the
    occurrence of the superradiant instability\cite{Detweiler:1980uk,Dolan:2007mj}.
Once the nonaxisymmetric boson cloud has grown, it dissipates through
gravitational waves on a much longer time scale
$\tau_{\rm GW}\sim 
5\times 10^{11}{\rm yr}\,
j_{\rm in}^{-1}
\left(10^6M_\odot/M\right)^{14}
\left(10^{-17}{\rm eV}/m_s c^2\right)
^{15}$.
Therefore we can assume that the superradiant instability occurs
quickly ($\tau_{\rm inst} \ll \tau_{\rm GW}$) and that the black
hole/cloud system is in equilibrium over a typical LISA observation time $T_{\rm obs}\sim 1{\rm yr}\ll \tau_{\rm inst} \ll \tau_{\rm GW}$.
The black hole/cloud system will remain in equilibrium even if there is accretion, because the (Salpeter) accretion timescale $\tau_{\rm acc}\gg T_{\rm obs}$\cite{Brito:2014wla}.

From an observational standpoint, there is no reason why three independent measurements of the boson mass using Eqs.~(\ref{eq:regge}), (\ref{eq:shape}) and (\ref{eq:evolutioneq}) should yield the same result, unless the superradiant instability hypothesis is correct. 
The gravitational waveform emitted by the EMRI of a small compact object orbiting the black hole/cloud system encodes both the host geometry and the gravitational potential of the cloud, making it possible to either confirm this hypothesis if the measurements are self-consistent or rule it out if they are not.
In other words, a measurement of either $\mu_s^{(i)}$ $(i=1,\,2,\,3)$ gives the boson mass only \emph{if} the boson cloud exists.  However, a self-consistent measurement of \emph{more than one} $\mu_s^{(i)}$ confirms the existence of the cloud.

EMRI observations by LISA can measure both the mass and spin of the host black hole to better than $1\%$ accuracy\cite{Babak:2017tow}. 
Matter effects may be resolved when the density of the surrounding material is sufficiently high: in fact, such matter effects are resolvable even when the density is much smaller than expected from boson clouds\cite{Eda:2013gg}. 
Therefore, as we show below, the tests we just outlined can be performed with LISA EMRI observations. 
For illustration: if the mass $M=10^5\,M_\odot$ and spin $a=0.6 M$ can
be measured to an accuracy $\sim$1\%, and $A$ and $B$ (with degeneracies) may be measured to
an accuracy $\sim$10\%, taking the 95\% confidence interval of $M_s \in [0,0.1 M]$.
Then the three estimates of the ultralight boson particle $\mu_s^{(1)}$, $\mu_s^{(2)}$ and $\mu_s^{(3)}$ would have errors $\sim$12\%, 8\% and 69\%, respectively.

As a proof of principle, let us first consider a case study of a system where
it is indeed possible to confirm the existence of ultralight bosons with LISA. 
We construct an EMRI gravitational-wave template in the black hole/cloud potential of Eq.~(\ref{eq:potential}), where $A$ and $B$ are free parameters. 
Following previous work\cite{Eda:2013gg}, to compute the evolution we include the lowest post-Newtonian (PN) order in the phasing as well as the leading order contribution from matter effects. 
We also add spin-dependent PN corrections to the \emph{inspiral}
waveform (as implemented in the LIGO Algorithm Library\cite{LAL}), which allows us to estimate the black hole spin\cite{Arun:2008kb, Bohe:2013cla}. 
Since the waveform includes matter effects, an EMRI observation allows us to infer {\em both} the boson cloud and host black hole properties: in particular, by matched filtering we can recover the masses and (aligned) spin of the central black hole, as well as the boson cloud amplitude and steepness parameters ($A$ and $B$).

\begin{figure}
\includegraphics[width=\columnwidth]{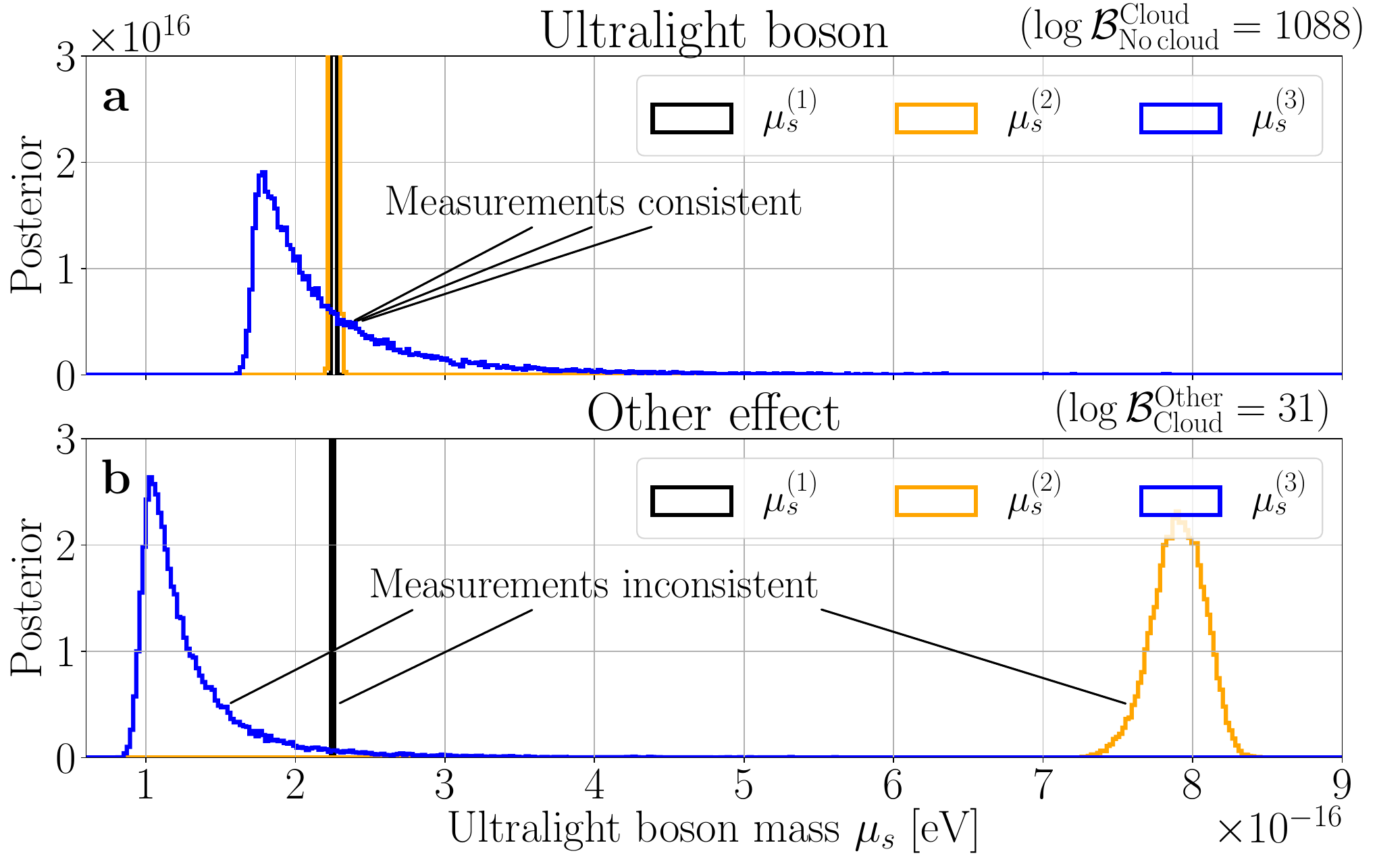}
 \caption{
 \textbf{Three independent posterior distribution measurements of ultralight boson particle masses $\mu_s^{(1)}$ (black), $\mu_s^{(2)}$ (orange) and $\mu_s^{(3)}$ (blue) [see Eqs.~(\ref{eq:regge})-(\ref{eq:evolutioneq})] from a single gravitational-wave observation with LISA.} 
 Panel \textbf{a}: the signal is produced by an EMRI into a black hole/cloud system with $\mu_s=2.26\times 10^{-16}$~eV. 
 All three measurements overlap with each other, favouring the presence of the cloud over the ``no cloud'' hypothesis with Bayes factors $\log B^{\text{Cloud}}_{\text{No cloud}} > 1000$ (top-right). 
 Panel \textbf{b}: the black hole has the same properties, but the ``cloud'' is produced by a dark matter mini-spike. 
 Measurements do not overlap, ruling out the boson cloud hypothesis with Bayes factors $\log B^{\text{Other}}_{\text{Cloud}}>30$.
 }
 \label{fig:axionmassinference}
\end{figure}

To be specific, we consider gravitational waves from a stellar-mass
black hole ($m=60$ $M_\odot$, $a^\prime = 0$) inspiralling into a
supermassive black hole ($M = 10^5$ $M_\odot$, $a=0.6 M$)
surrounded by a cloud generated by bosons of mass $\mu_s=2.26\times
10^{-16}$~eV, with total cloud mass $M_s=0.05 M$, one year
observation time and a LISA signal-to-noise ratio (SNR) $(h,h)^{1/2}=97$, which corresponds to redshift $z\sim1$\cite{Gair:2004iv,Babak:2017tow}.
We use a nested sampling Markov-Chain Monte Carlo algorithm to recover three independent posteriors $\mu_s^{(i)}$ $(i=1,\,2,\,3)$ from measurements of $M$, $a$, $A$ and $B$. 
Figure~\ref{fig:axionmassinference} (top panel) shows that in this case we confirm the ultralight boson hypothesis because all three measurements overlap.
In the bottom panel of Figure~\ref{fig:axionmassinference} we consider instead the gravitational wave signal produced by a small compact object falling into a black hole surrounded by a dark matter mini-spike with $\rho_{\text{sp}}=3\times10^5 M_\odot/\text{AU}^3$, $\alpha=1$ and $r_{\text{sp}}=6 M$ (see "Methods" for the motivation for the parameters)\cite{Eda:2013gg,Eda:2014kra}. 
In this case we can rule out ultralight bosons as a source of the matter distribution, because the recovered ultralight boson masses do not overlap.

\begin{figure}
 \includegraphics[width=\columnwidth]{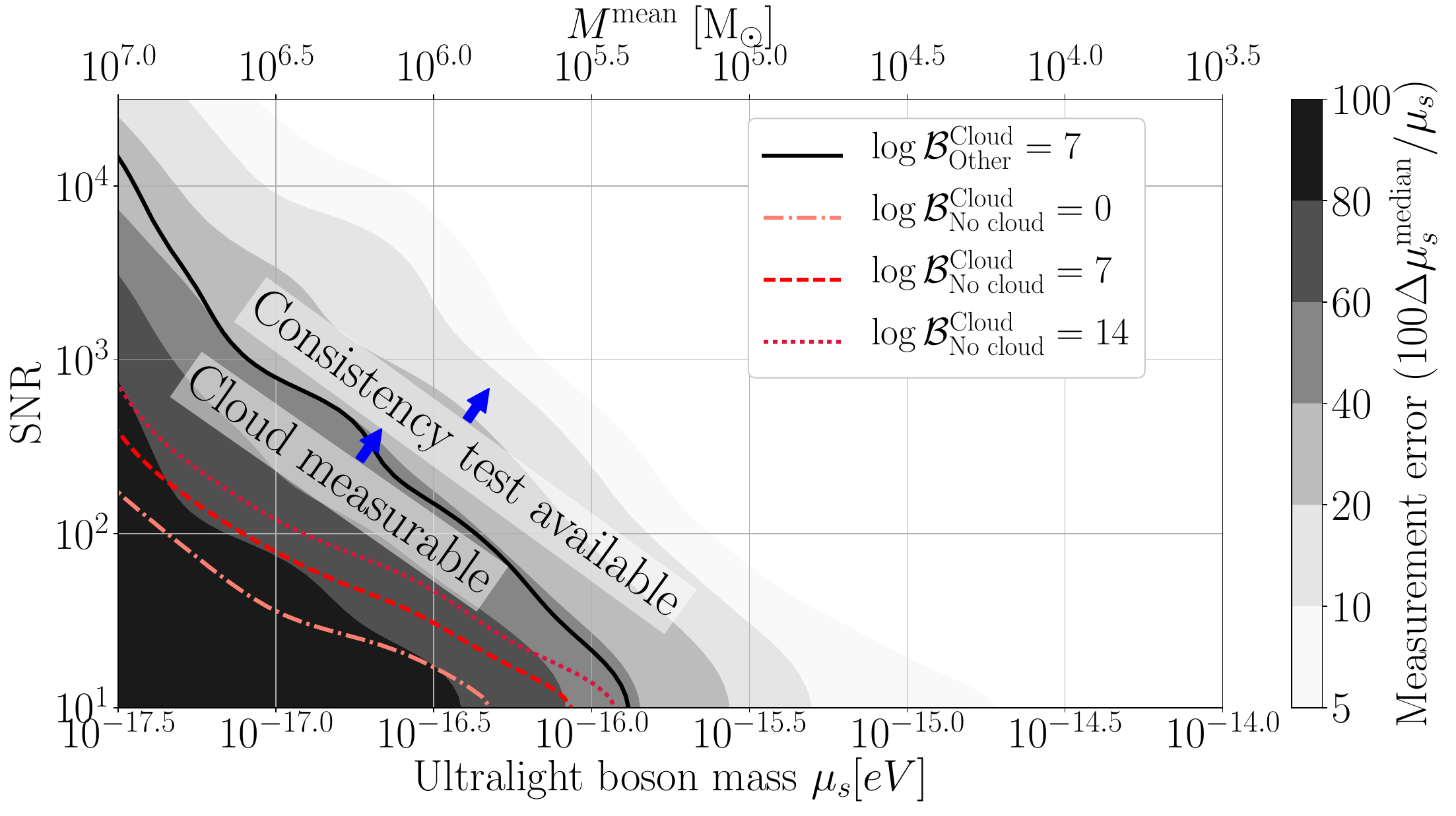}
 \caption{
 \textbf{Median error of the measurements $\mu_s^{(1,2,3)}$ in color as a function of the SNR and ultralight boson mass $\mu_s$ and mean host black hole mass $M^\text{mean}$.} 
  The consistency test is available across a wide $\mu_s \in  [10^{-16.5},10^{-14}]$ eV (i.e. above the black contour, $\log  \mathcal{B}^{\text{Cloud}}_{\text{Other}}=7$). The presence of the  cloud can be inferred over $\mu_s \in [10^{-17},10^{-14}]$ eV (above  the red dashed contour, $\log \mathcal{B}^{\text{Cloud}}_{\text{No cloud}}=7$).  We have simulated a population of binaries distributed logarithmically across host mass $M \in [10^3, 10^7] \rm M_\odot$ and mass ratio $q\in [10^{-3}, 10^{-2}]$, and linearly in host spin $j\in [0.4,0.98]$. 
  We average all measurements over spin, and use a Gaussian filter on
  the data for visualization purposes.
    The black solid contour corresponds to $\log
    \mathcal{B}^{\text{Cloud}}_{\text{Other}}=7$; the red
    dash-dotted, dashed, and pink dotted contours correspond to $\log
    \mathcal{B}^{\text{Cloud}}_{\text{No Cloud}}=0$, $7$ and $14$, respectively. 
 }
 \label{fig:summary}
\end{figure}

It is natural to ask whether the case study shown above is generic:
can similar measurements be done across a range of binary parameters and boson masses?
To answer this question we simulated one year of LISA EMRI
observations for different boson masses by varying the host spin
$j\in[0.4,0.98]$, mass $M$, and SNR. To sample this large
parameter space we used a (much faster) Fisher information matrix
calculation to recover parameters.  
As shown in Figure~\ref{fig:summary},  we found that it is possible to carry out consistency checks for two
or more of the $\mu_s^{(i)}$'s for a broad range of binary parameters
and for boson masses $\mu_s \in [10^{-17}, 10^{-14}]$ eV. 
Extending the observation time improves the lower bound, but masses
$\mu_s \lesssim 10^{-20}\,\text{eV}$ result in unmeasurable effects (the waveform  phase correction due to the cloud contributes less
than one gravitational-wave cycle).
At the opposite end of the mass range, black hole/cloud systems with
$\mu_s \gtrsim 10^{-14}\,\text{eV}$ would produce EMRI signals outside
of the LISA sensitivity band. 

Our work is meant to be a proof-of-principle demonstration that binary
pulsar-like tests of ultralight bosons are possible through EMRI
observations with LISA, 
but future efforts to model the system more precisely will be required 
for the practical implementation of this program in the LISA data-analysis. 
Gravitational waves from EMRIs can be computed to high accuracy for astrophysical massive black holes in isolation, which are characterized only by their mass and spin, but surrounding matter can back-react on the binary. 
Back-reaction effects are small for the high mass ratios considered here, but they can be relevant for comparable-mass binaries.

Furthermore, it will be interesting to take into account the possibility of "mode mixing": perturbations due to the small orbiting companion can mix superradiating modes with ``dumping'' (infalling) levels of the cloud, causing the cloud to collapse before the binary can trace its properties\cite{Arvanitaki:2014wva}. 
If this occurs, our test will yield a null result with non-overlapping distributions for the $\mu^{(i)}$'s, as we would not measure the effects of the cloud. 
The disturbance on the cloud due to a companion is an active area of research: recent work suggests that the boson cloud would, in fact, survive mode mixing in the high mass ratio scenario studied in this paper when the small object is in a corotating orbit\cite{Baumann:2018vus}, and then a measurement would be possible.
Counter-rotating EMRIs are less likely to be detectable by LISA\cite{2018LRR....21....4A}. 
Numerical estimates of the cloud mass after depletion can be approximated by the phenomenological fits $\sim M_s (1- \exp(-10 q^{1.2} (M \mu_s)^{0.18}))$ and $\sim M_s (1-\exp(-2.3 q^{0.9}-10 (M \mu_s)^{3/2}))$ for co-rotating and counter-rotating orbits, respectively, as long as $q \lesssim 0.1$ and $M \mu_s\lesssim 0.2$ [Ref.~\citen{Baumann:2018vus}]. Cloud depletion can therefore be neglected for most EMRIs, except for a narrow region of parameter space involving astrophysically rare counter-rotating orbits.

We have accounted for possible degeneracies between the binary parameters and effects due to the boson cloud using nested sampling to simultaneously infer the binary's intrinsic parameters (masses, spins), the cloud properties and the merger time (maximizing over the phase of the wave). 
The effect of the cloud is a slow cumulative shift uniquely tied to the cloud's density profile, so (as expected) we found that the degeneracy between the orbital parameters of the binary and the cloud parameters is small. 
Figure~\ref{fig:axionmassinference} shows that the consistency test can be performed for SNRs that can be achieved with LISA\cite{Babak:2017tow}.
We have also verified that the consistency test is possible when including eccentricity corrections at first order in the gravitational-wave phasing\cite{PhysRevD.94.064020}. 

Following previous work\cite{Brito:2014wla,Ferreira:2017pth} we focus on the most unstable mode, as it accounts for most of the matter distribution.  Higher modes are unlikely to be observed because they would only become unstable on much longer timescales ($\tau_{\rm  inst}^{\rm HM}\sim10^7$ years for the illustrative case considered here)\cite{Dolan:2007mj}.  This is appropriate for the present order-of-magnitude estimate of the effect of the matter distribution.  However, the next-to-leading order mode can be filled through superradiance on a time scale one or two orders of magnitude higher than the first mode's dissipation time-scale. If the fundamental mode is depleted via gravitational radiation, our proposed test can still be performed by replacing Eqs.~(\ref{eq:regge})-(\ref{eq:evolutioneq}) with the equivalent expression for the next-to-leading order mode.

For simplicity, we assumed that the small compact object is in an equatorial orbit and we computed the gravitational potential for real scalar fields. 
However our results also apply to complex scalar fields where the potential is stationary\cite{Herdeiro:2014goa}.
In the same spirit we used PN waveforms with aligned spins, as opposed to more realistic, fully precessing EMRI waveforms with eccentricity\cite{Babak:2006uv,Chua:2017ujo,Babak:2017tow}.
These corrections will matter in LISA data analysis, but they only contribute a fraction of the total phase shift, and so they can be omitted for order-of-magnitude estimates.
We have also checked the convergence of the PN expansion by comparing the accumulated phase shift of the highest term relative to the next-to-highest term, finding the difference to be negligible at the percent level.
Another interesting effect is that, because the potential of a boson cloud is not spherically symmetric, orbital resonances could result in angular momentum transfer between the companion and the cloud, increasing the orbital eccentricity\cite{Ferreira:2017pth}.
These resonances are an interesting topic for future study, but we verified that they do not occur for the orbital parameters considered here.

In conclusion, the possibility to obtain three independent measurements of ultralight boson masses which can be cross-compared for consistency is mostly unaffected by the corrections listed above.
We have demonstrated that LISA EMRI observations, in a realistic SNR range, can be used to confirm (or rule out) the formation of ultralight boson condensates around astrophysical black holes in the mass range $\mu_s \in [10^{-16.5},10^{-14}] \, \rm eV$, directly probing the existence of ultralight bosons.  
More accurate waveform models and more accurate treatments of superradiance (including higher-order modes and possible transitions among superradiant states) will be needed for an implementation of this idea in LISA data analysis.

\bibliographystyle{naturemag}

\begin{addendum}

\item[Acknowledgements] We thank Caio Macedo, Jo\~ao Luis Rosa and Guilherme Raposo for useful discussions on cloud depletion.
O.A.H. is supported by the Hong Kong Ph.D. Fellowship Scheme (HKPFS) issued by the Research Grants Council (RGC) of Hong Kong. 
E.B. and K.W. are supported by NSF Grant No. PHY-1841464, NSF Grant No. AST-1841358, NSF-XSEDE Grant No. PHY-090003, and NASA ATP Grant No. 17-ATP17-0225. 
R.B. acknowledges financial support from the European Union's Horizon 2020 research and innovation programme under the Marie Sk\l odowska-Curie grant agreement No. 792862.
This project has received funding from the European Union's Horizon 2020 research and innovation programme under the Sk\l odowska-Curie grant agreement No. 690904.
T.G.F.L. was partially supported by grants from the Research Grants Council of the Hong Kong (Project No. CUHK14310816 and CUHK24304317) and the Direct Grant for Research from the Research Committee of the Chinese University of Hong Kong.
The authors would like to acknowledge networking support by the
GWverse COST Action CA16104, ``Black holes, gravitational waves and fundamental physics.''

\item[Author Contributions] 
O.A.H. conceived the idea of probing the existence of ultralight bosons, performed the full analysis, led the project and wrote the initial manuscript. 
K.W. surveyed the full LISA parameter space for the measurement and performed key analysis for the second figure.
T.G.F.L. closely supervised all parts of the project, including its implementation and writing of the manuscript, and played an instrumental role in the development of the idea. 
E.B. made major revisions and significant contributions to the manuscript and provided theoretical input on superradiance and the overall study. 
R.B. provided detailed theoretical insight on superradiance, its coupling to gravitational waves, pointed out an error in the initial study that contributed to the final results, and contributed significantly to the manuscript. 
All authors commented on the manuscript and wrote parts of it.

\item[Correspondence and requests for materials] should be addressed to O.A.H. (email: hannuksela@phy.cuhk.edu.hk).

\item[Competing Interests] 
Authors declare no competing interests.

\end{addendum}

\clearpage

\clearpage

\section*{Methods}
\noindent
\textbf{Posterior estimation.} We consider a LISA EMRI signal from a black hole/cloud system and use Nested Sampling (as implemented in \texttt{MultiNest}\cite{Feroz:2007kg,Feroz:2008xx,Feroz:2013hea}) to evaluate the posterior distribution of our measurement, with the likelihood defined for colored gaussian noise following the LISA power spectral density (PSD)\cite{Cornish:2018dyw}
\begin{equation} \label{eq:loglikelihood}
 \log \mathcal{L}(\vec{\theta}, A, B) = (s,h(\vec{\theta}, A, B))-\frac{1}{2}(h(\vec{\theta}, A, B),h(\vec{\theta}, A, B)).
\end{equation}
Here $s$ is the injected signal, which we assume to be noiseless (this is approximately true at high SNR, as in our chosen scenario) and $h(\vec{\theta}, A, B)$ is the gravitational-wave template at 3.5 PN order\cite{Arun:2008kb,Mikoczi:2005dn,Bohe:2013cla} for a detector oriented optimally for the plus polarized wave, which we use in our parameter estimation to sample over both binary parameters $\vec{\theta}$ and matter parameters $A,\, B$.

The gravitational potential due to the cloud is included at lowest PN order\cite{Eda:2013gg}:
\begin{equation}\label{eq:waveform}
 h(f) = h(f) e^{i (\psi(f) + \Delta \psi_{\text{matter}}(f) + 2 \pi f t_c + 2 \phi_c)},
\end{equation}
where $h(f)$, $\psi(f)$ are the amplitude and phase of the gravitational wave, $\Delta \psi_{\text{matter}}(f)$ is the phase shift due to matter effects, and ($t_c$, $\phi_c$) are the time and phase of coalescence.

The inner product $(a,b)$ is defined as
\begin{equation}
 (a,b) = 4 \Re \left[ \int_0^{\infty} \frac{a(f) b^*(f)}{S_n(f)} df \right], 
\end{equation}
where $S_n(f)$ is the LISA PSD\cite{Cornish:2018dyw}. 
We take the absolute value of the inner product to maximize over the phase of coalescence\cite{Jaranowski:1998qm,Allen:2005fk}.
In our parameter estimation, we simulate a single gravitational-wave event with given parameters and sample over binary masses ($m_1,\,m_2$), aligned spins ($s_1,\,s_2$), time of coalescence $t_c$, and boson cloud parameters ($A,\,B$): see Eq.~(\ref{eq:potential}).

Since the nested sampling approach is computationally expensive, we use the Fisher information matrix approach\cite{Berti:2004bd} to explore the full parameter space (Figure~\ref{fig:summary}).
The elements of the Fisher matrix are calculated by taking the inner product between derivatives of waveform 
\begin{equation}
\Gamma_{ab} \equiv \left( \frac{\partial h}{\partial {\theta}^{a}}, \frac{\partial h}{\partial {\theta}^{b}} \right).
\end{equation}
The measurement uncertainties and correlations are given by
\begin{equation}
\begin{split}
\Delta{\theta}^{a} &= \sqrt{{\Sigma}^{aa}}, \\
{c}_{ab} &= \frac{{\Sigma}^{ab}}{\sqrt{{\Sigma}^{aa}{\Sigma}^{bb}}}, \\
\end{split}
\end{equation}
respectively, where $\Sigma^{ab} = \Gamma_{ab}^{-1}$.
In this approach, we implemented the gravitational-wave template at 2PN order\cite{Berti:2004bd}.
We have verified that the Fisher matrix estimates are consistent with the nested sampling results.

\textbf{Bayes factors.} 
We define three hypotheses: (1) there is a matter cloud produced by ultralight bosons ($\mathcal{H}_{\text{Cloud}}$); (2) there is some matter distribution similar to a bosonic cloud, but not necessarily produced by ultralight bosons: i.e., Eqs.~(\ref{eq:regge})-(\ref{eq:evolutioneq}) are not necessarily satisfied ($\mathcal{H}_{\text{Other}}$); and (3) there is no cloud ($\mathcal{H}_{\text{No cloud}}$). 
The corresponding evidences are:
\begin{align}
\nonumber
  \mathcal{Z}_{\text{Cloud}}         &= 
    \int \mathcal{L}(\vec{\theta}, f(M,a,M_s), g(M,a)) \pi(\vec{\theta}) \pi(M_s) d\vec{\theta} dM_s, \\  
\nonumber
  \mathcal{Z}_{\text{Other}}         &= 
 \int \mathcal{L}(\vec{\theta}, A, B) \pi(\vec{\theta}) \pi(A) \pi(B) d\vec{\theta} dA dB, \\
 \mathcal{Z}_{\text{No cloud}}      &= 
 \int \mathcal{L}(\vec{\theta}, 0, 0) \pi(\vec{\theta}) d\vec{\theta}, 
\end{align}
with 
\begin{align}\label{eq:ABConstraints}
\nonumber
  f(M,a,M_s) &= \frac{a^2}{4 \sqrt{2 \pi}\left(1+\alpha\right)}\sqrt{ \frac{M_s}{32 M^5 \left(1+\alpha\right)-17 a^2 M^3} },\\
  g(M,a)     &= \frac{a^2}{4 M^3 \left(1+\alpha\right)^2},
\end{align}
where $\alpha\equiv\sqrt{1-(a/M)^2}$, $\vec{\theta}$ are the binary parameters, $\mathcal{L}(\vec{\theta}, A, B)$ is the likelihood, and $f(M,a,M_s)$ and $g(M,a)$ can be found by solving Eqs.~(\ref{eq:regge})-(\ref{eq:evolutioneq}) for $A$ and $B$.
We computed the Bayes factor of two hypotheses $\mathcal{H}_x$ and $\mathcal{H}_y$, i.e. the ratio
$\mathcal{B}^{y}_x = \mathcal{Z}_y/\mathcal{Z}_x$,
using nested sampling. 
Since alternative hypotheses typically cause only small corrections to the gravitational waveform, we set conservatively a uniform prior on $(A,\,B)$ so that the effect on the waveform phase is at most of order percent, but still measurable ($\gtrsim 1$ rad). 
We additionally cut off the prior at a maximum of $(3A,\, 3B)$, to be conservative. 
Setting a completely uniform prior or allowing for negligible phase deviations would yield more optimistic results. 
$M_s$ is set to be uniform over $M_s \in [0, 0.1] M$. 
When using the Fisher information matrix, we compute the Bayes factors following standard methods\cite{2012PhRvD..86h2001V}. We have verified that the results are compatible with the nested sampling analysis.

\textbf{Waveform.} We estimate the waveform by perturbing the energy balance equation\cite{Eda:2013gg}. 
The gravitational potential produced by the boson cloud is\cite{Ferreira:2017pth}
\begin{equation}\label{eq:potential} 
 \begin{split}
  \Phi_b(r) &\simeq \left[ \pi  A^2  e^{-B r} \left(-M B^6  r^5-2 B^5 r^4 (M-2 r) \right. \right. \\
  &\left. \left. -12 B^4 r^3 (M-2 r)+8 B^3 r^2 (10 r-3 M) \right. \right. \\
  &\left. \left. -16 B^2 r (M-10 r) +16 e^{B r} \left(B^3 M r^2-4 B^2 r^2 \right. \right. \right. \\
  &\left. \left. \left. +B M-12\right)-16 B (M-12 r)+192\right)\right]/ \left(2 B^4 M r^3\right).
 \end{split}
\end{equation}
We expand the phase shift to first order in $\Phi_b/\Phi_{\text{BH}}$, where $\Phi_{\text{BH}}$ is the gravitational potential in the absence of the cloud. 
This introduces a correction to Kepler's law and the energy balance and changes the accumulated orbital phase shift. 
The phase shift due to matter can be computed from the orbital energy balance equation\cite{Eda:2013gg}
\begin{equation} \label{eq:energybalance}
 \frac{dE_{\rm orbit}}{dt} + \frac{dE_{\rm gw}}{dt} = 0,
\end{equation}
where
\begin{equation}
\begin{split}
 \frac{dE_{\rm orbit}}{dt} &= \frac{d}{dt} \left( \frac{1}{2} \mu  v^2+\mu\Phi (r) \right),\\
 \frac{dE_{\rm gw}}{dt}    &= \frac{32 \mu^2 r^4 \omega^6}{5},
\end{split}
\end{equation}
and $\Phi(r)=-M/r+\Phi_b(r)$.
Here $\mu$ is the small compact object mass, $v$ is the velocity of the companion, and $\omega$ is the orbital angular frequency. 
This gives the rate of change of the orbital radius
\begin{equation}
 r'(t) = -\frac{32 \mu ^2 r^4 \omega^6}{5 \left[\mu  v v'(r)+\mu\Phi'(r)\right]},
\end{equation}
which can be translated to the total gravitational-wave phase shift using the stationary phase approximation\cite{Maggiore:1900zz}:
\begin{equation}
 \Delta \psi_{\text{matter}} = 2 \pi f t(f) - 2 \phi(f),
\end{equation}
where the time and orbital phase are given by
\begin{equation}
\begin{split}
 t(f)    &= \int \frac{1}{r'(t)} dr,\\
 \phi(f) &= \int \frac{\omega}{r'(t)} dr.
\end{split}
\end{equation}
where we only consider matter contribution. 
The mapping between the orbital radius and the gravitational-wave frequency may be solved by inverting the following relation for $r$
\begin{equation}
 \omega(f) = \pi f = \sqrt{\frac{\Phi '(r)}{\mu r}},
\end{equation}
and expanding to first order in $\epsilon=\Phi_b/\Phi_{\text{BH}}$\cite{Eda:2013gg}. 
We have included first-order corrections from matter effects, and verified that second-order corrections cause negligible phase corrections in the gravitational waveform at percent level in the case that we consider here.

\textbf{Dark matter mini-spike.} To illustrate a case where our black hole/cloud test discriminates other effects from boson clouds, we consider a black hole surrounded by a dark matter mini-spike (similar constructions were used in previous work\cite{Eda:2013gg}). 
We assume the mini-spike density to follow a power-law
\begin{equation}
 \rho(r) = \rho_{\text{sp}} \left( \frac{r}{r_{\text{sp}}} \right)^{-\alpha},
\end{equation}
where $\rho_{\text{sp}}$ and $r_{\text{sp}}$ are the density and radius normalization constants, and $\alpha$ gives the steepness of the profile. 
We follow\cite{Eda:2013gg} to construct the orbital phase shift of the gravitational wave due to the mini-spike. 
To assess whether the dark matter mini-spike can mimic a boson cloud we first construct the orbital phase shift due to a dark matter spike, then we fit the results using a boson cloud, but treating the $A$ and $B$ parameters as free. 
We set $\rho_{\text{sp}}=3\times10^5 M_\odot/\text{AU}^3$, $\alpha=1$ and $r_{\text{sp}}=6 M$, which causes an orbital shift of similar order as the boson cloud in our example scenario. 
 
We chose the dark matter mini-spike parameters to mimic boson cloud effects. 
If we had chosen the dark matter mini-spike profile expected to form through adiabatic growth from a seed black hole in a typical cuspy dark matter environment with density $\sim \text{GeV}/\text{cm}^3$ at $100$ kpc [Ref.~\citen{Gondolo:1999ef}], or if we had chosen different values of $\alpha$, the discriminatory power of our test would have improved even further.

\begin{addendum}

\item[Data availability statement] 
The data that supports the plots within this paper and other findings of this study are available from the corresponding author upon reasonable request.

\end{addendum}

\end{document}